\documentclass[a4paper]{jpconf}
\usepackage{graphicx}

\begin{document}
\title{Exploring the partonic phase at finite
chemical potential in and out-of equilibrium}

\author{ O. Soloveva $^{1}$, P. Moreau $^{2}$,  L. Oliva $^{1}$,
 T. Song $^{3}$, Cassing $^{4}$ and E.~Bratkovskaya $^{1,3}$,}

\address{$^1$ Institute for Theoretical Physics, University of Frankfurt, Frankfurt, Germany}

\address{$^2$ Department of Physics, Duke University, Durham, NC 27708, USA}

\address{$^3$ Institute for Theoretical Physics, University of Giessen, Giessen, Germany}

\address{$^3$ GSI Helmholtzzentrum f\"{u}r Schwerionenforschung GmbH,  Darmstadt, Germany}


\ead{E.Bratkovskaya@gsi.de}

\begin{abstract}
We study the influence of the baryon chemical potential $\mu_B$  on the properties
of the Quark--Gluon--Plasma (QGP) in and out-of equilibrium. The description of the QGP in equilibrium is based on the effective propagators and couplings from the Dynamical QuasiParticle Model (DQPM) that is matched to reproduce the equation-of-state of the partonic system above the deconfinement temperature $T_c$ from lattice Quantum Chromodynamics (QCD).
We calculate the transport coefficients such as the ratio of shear viscosity $\eta$ and  bulk viscosity $\zeta$ over entropy density $s$, i.e., $\eta/s$ and $\zeta/s$ in the $(T,\mu_B)$ plane and compare to other model results available at $\mu_B =0$.
The out-of equilibrium study of the QGP is performed within
the Parton--Hadron--String Dynamics (PHSD) transport approach extended in the
partonic sector by explicitly calculating the total and differential partonic
scattering cross sections (based on the DQPM propagators and couplings) evaluated
at the actual temperature $T$ and baryon chemical potential $\mu_B$ in each
individual space-time cell of the partonic scattering.
The traces of their $\mu_B$ dependences are investigated in
different observables for relativistic heavy-ion collisions with a focus on the  directed and elliptic flow
coefficients $v_1, v_2$ in the energy range 7.7 GeV $\le \sqrt{s_{NN}}\le 200$ GeV.
\end{abstract}



\section{Introduction}
The phase diagram of matter is one of the most fascinating subjects in physics, which
also has important implications on chemistry and biology. Its phase boundaries and (possibly) critical points have been the focus of physics research for centuries.
Apart from the traditional phase diagram in the plane of temperature $T$ and pressure $P$,
its transport properties like the shear and bulk viscosities, the electric conductivity, etc., are also of fundamental interest. These transport coefficients emerge from the stationary limit of correlators and provide additional information on the systems in thermal and chemical equilibrium apart from the equation of state (EoS).
In particular the phase diagram (PD) of strongly interacting matter has been the topic of interest for the last decades and substantial experimental and theoretical efforts have been invested to shed light on this issue. The PD contains the information about the properties of our universe from the early beginning---directly after the Big Bang---when the matter was in the quark-gluon plasma (QGP) phase at very high temperature $T$ and practically zero baryon chemical potential $\mu_B$,
to the later stages of the universe, where in the expansion phase stars and Galaxies have been formed. Here, the matter is at low temperature, however, large baryon chemical potential.
Relativistic and ultra-relativistic heavy-ion collisions (HICs) nowadays offer the unique possibility to study some of these phases, in particular a QGP phase and its phase
boundary to the hadronic one. We point out that the phase diagram of strongly interacting matter in the $(T, \mu_B$) plane can also be explored in the astrophysical context at moderate temperatures and high $\mu_B$ \cite{Klahn:2006ir}, i.e. in the dynamics of supernovae or---more recently---in the dynamics of neutron-star merges.

In case of relativistic heavy-ion collisions one generates hot
and dense matter in the laboratory although within small space-time
regions. Whereas in low energy collisions one produces dense nuclear
matter with moderate temperature $T$ and large baryon chemical potential
$\mu_B$, ultra-relativistic collisions at Relativistic Heavy Ion
Collider (RHIC) or Large Hadron Collider (LHC) energies produce
extremely hot matter at small baryon chemical potential. In order to
explore the phase diagram of strongly interacting matter as a
function of $T$ and $\mu_B$ both type of collisions are mandatory.
According to lattice calculations of quantum chromodynamics
(lQCD)~\cite{Bernard:2004je,Aoki:2006we,Aoki:2009sc,Bazavov:2011nk}, the phase
transition from hadronic to partonic degrees of freedom (at
vanishing baryon chemical potential $\mu_B$=0) is a crossover. This
transition is expected to turn into a first order transition
at some critical point $(T_r, \mu_{cr})$ in the phase diagram with
increasing baryon chemical potential $\mu_B$. Since this critical
point cannot be  determined theoretically in a reliable way the beam
energy scan (BES) program  at  RHIC  aims to find the
critical point and the phase boundary by gradually decreasing the
collision energy~\cite{Mohanty:2011nm,Kumar:2011us}. Furthermore,
new facilities such as FAIR (Facility for Antiproton and Ion
Research) and NICA (Nuclotron-based Ion Collider  fAcility) are
under construction to explore in particular the intermediate energy
regime where one might study also the competition between chiral
symmetry restoration and deconfinement as suggested in Refs.
\cite{Cas16,Palmese}.

Current methods to explore QCD in Minkowski space for non-vanishing
quark (or baryon) densities (or chemical potential) are effective approaches.
Using effective models, one can
study the properties of QCD in equilibrium, i.e.,~thermodynamic quantities
as well as transport coefficients. To this aim, the dynamical quasiparticle model (DQPM)
has been introduced \cite{Peshier:2005pp,Cassing:2007nb,Cassing:2007yg,Linnyk:2015rco,Hamsa:JModPhys16},
which is based on partonic propagators with sizeable
imaginary parts of the self-energies incorporated. Whereas the real part of
the self-energies can be attributed to a dynamically generated mass (squared), the
imaginary parts contain the information about the interaction rates of the degrees-of-freedom.
Furthermore, the imaginary parts of the propagators define the spectral functions of the 'particles' which might show narrow (or broad) quasiparticle peaks.
A further advantage of a propagator based approach is that one can formulate a
consistent thermodynamics~\cite{Baym} as well as a causal theory for non-equilibrium configurations on the basis of Kadanoff--Baym equations~\cite{KadanoffBaym}.

Since relativistic heavy-ion collisions start with impinging nuclei in their groundstates, a proper non-equilibrium description of the entire dynamics through possibly different phases up to the final asymptotic hadronic states---eventually showing some degree of equilibration---is mandatory. To this aim, the Parton--Hadron--String Dynamics (PHSD) transport approach
~\cite{Linnyk:2015rco,Cassing:2008sv,Cassing:2008nn,Cassing:2009vt,Bratkovskaya:2011wp}
has been formulated more then a decade ago (on the basis of the
Hadron-String-Dynamics (HSD) approach~\cite{Cassing:1999es}),
and it was found to well describe observables from p+A and A+A collisions from SPS to LHC energies including electromagnetic probes such as photons and dileptons
as well as open cham hadrons \cite{Linnyk:2015rco,Linnyk:2013,Song:2018}.
In order to explore the partonic systems at higher $\mu_B$, the PHSD approach has been recently extended to incorporate partonic quasiparticles and their differential cross sections that depend not only on temperature $T$ as in the previous PHSD studies, but
also on  chemical potential $\mu_B$ explicitly \cite{Moreau:2019vhw}.
Within this extended approach we have previously studied the `bulk' observables  in HICs from AGS to RHIC energies for symmetric and asymmetric systems. However, we have found only a small influence of $\mu_B$ dependences of parton properties (masses and widths) and their interaction
cross sections in the bulk observables~\cite{Moreau:2019vhw}.

In this contribution we extend our study with respect to  the
collective flow ($v_1$, $v_2$) coefficients for different identified hadrons and their sensitivity to the $\mu_B$ dependences of partonic cross sections. 

\section{The PHSD Approach}

We start with recalling the basic ideas of the PHSD transport approach and the dynamical quasiparticle model (DQPM).
The~Parton--Hadron--String Dynamics (PHSD) transport approach \cite{Linnyk:2015rco,Cassing:2008sv,Cassing:2008nn,Cassing:2009vt,Bratkovskaya:2011wp}
is a microscopic off-shell transport approach for the description of strongly interacting hadronic and partonic matter in and out-of equilibrium. It is based on the solution of
Kadanoff--Baym equations in first-order gradient \mbox{expansion
\cite{Cassing:2008nn}} employing `resummed' propagators from the DQPM \cite{Peshier:2005pp,Cassing:2007nb,Cassing:2007yg}  for the partonic~phase.
The DQPM, furthermore, has been introduced in Refs.~\cite{Peshier:2005pp,Cassing:2007nb,Cassing:2007yg} for the effective description
 of the QGP in terms of strongly interacting quarks and gluons
with properties and interactions which
are adjusted to reproduce lQCD results for 
 the equilibrated QGP at finite temperature $T$ and
baryon (or quark) chemical potential $\mu_B$.
In the DQPM, the quasiparticles are characterized by single-particle Green's functions (in propagator representation) with complex self-energies.
The real part of the self-energies is related to the dynamically generated parton masses (squared),
whereas the imaginary part provides information about the lifetime and/or
reaction rates of the degrees-of-freedom.

In PHSD, the partons (quarks and gluons) are  characterized by broad spectral functions $\rho_j$ ($j=q, {\bar q}, g$) and thus  are off-shell contrary to the conventional cascade or transport models
dealing with on-shell particles, i.e.,~the $\delta$-functions in the invariant mass squared.
The quasiparticle spectral functions are assumed to have a Lorentzian form~\cite{Linnyk:2015rco}, which are specified by 
 the parton masses and width parameters:
\begin{eqnarray}
\!\!\!\!\!\! \rho_j(\omega,{\bf p}) =
 \frac{\gamma_j}{E_j} \left(
   \frac{1}{(\omega-E_j)^2+\gamma_j^2} - \frac{1}{(\omega+E_j)^2+\gamma_j^2}
 \right)\
\label{eq:rho}
\end{eqnarray}
separately for quarks/antiquarks and gluons ($j=q,\bar{q},g$). With the convention $E^2({\bf p}^2) = {\bf p}^2+M_j^2-\gamma_j^2$, the parameters $M_j^2$ and $\gamma_j$ are directly related to the real
and imaginary parts of the retarded self-energy, e.g.,~$\Pi_j =
M_j^2-2i\gamma_j\omega$.

\begin{figure}[h]
\centerline{ \includegraphics[width=75mm]{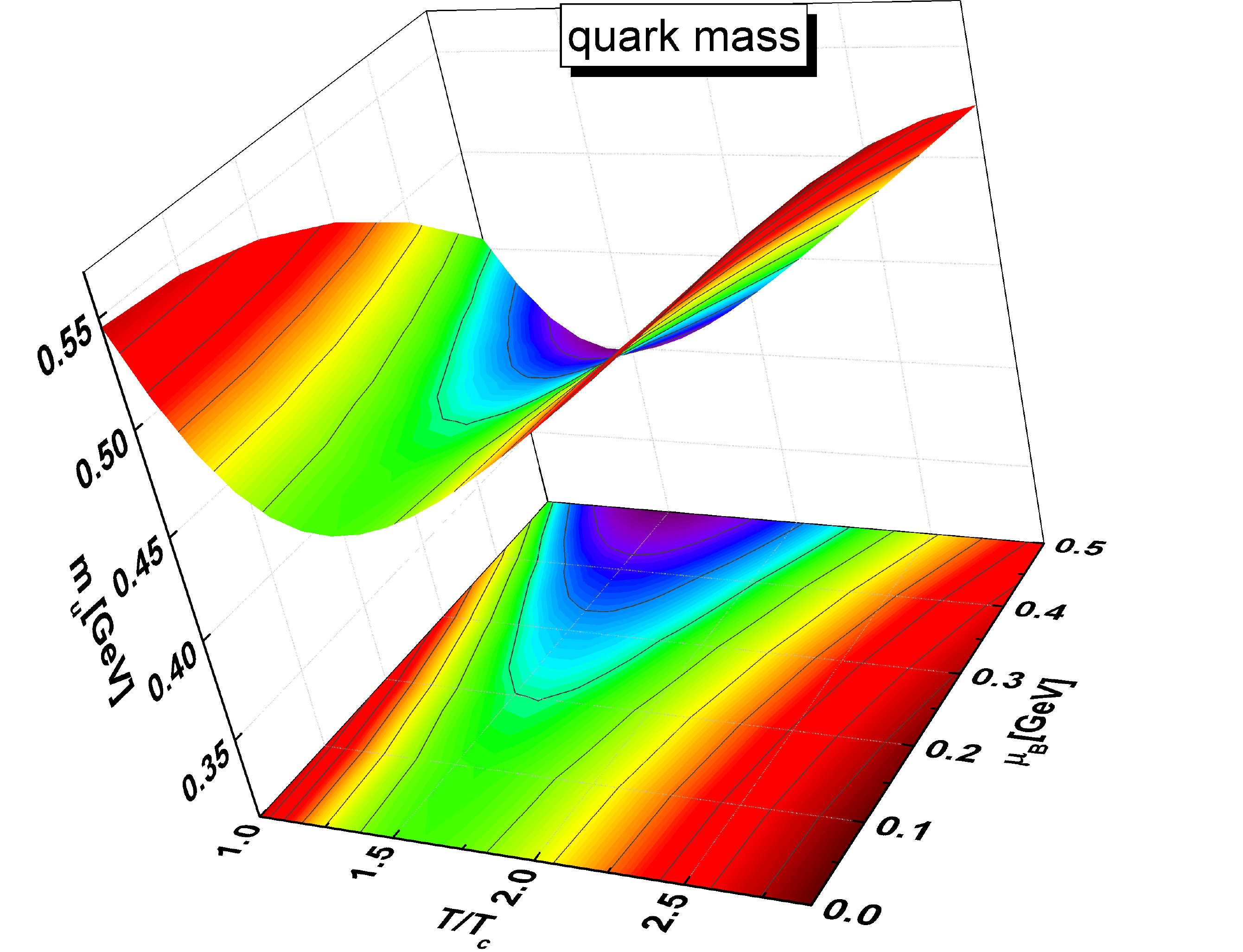}
\includegraphics[width=75mm]{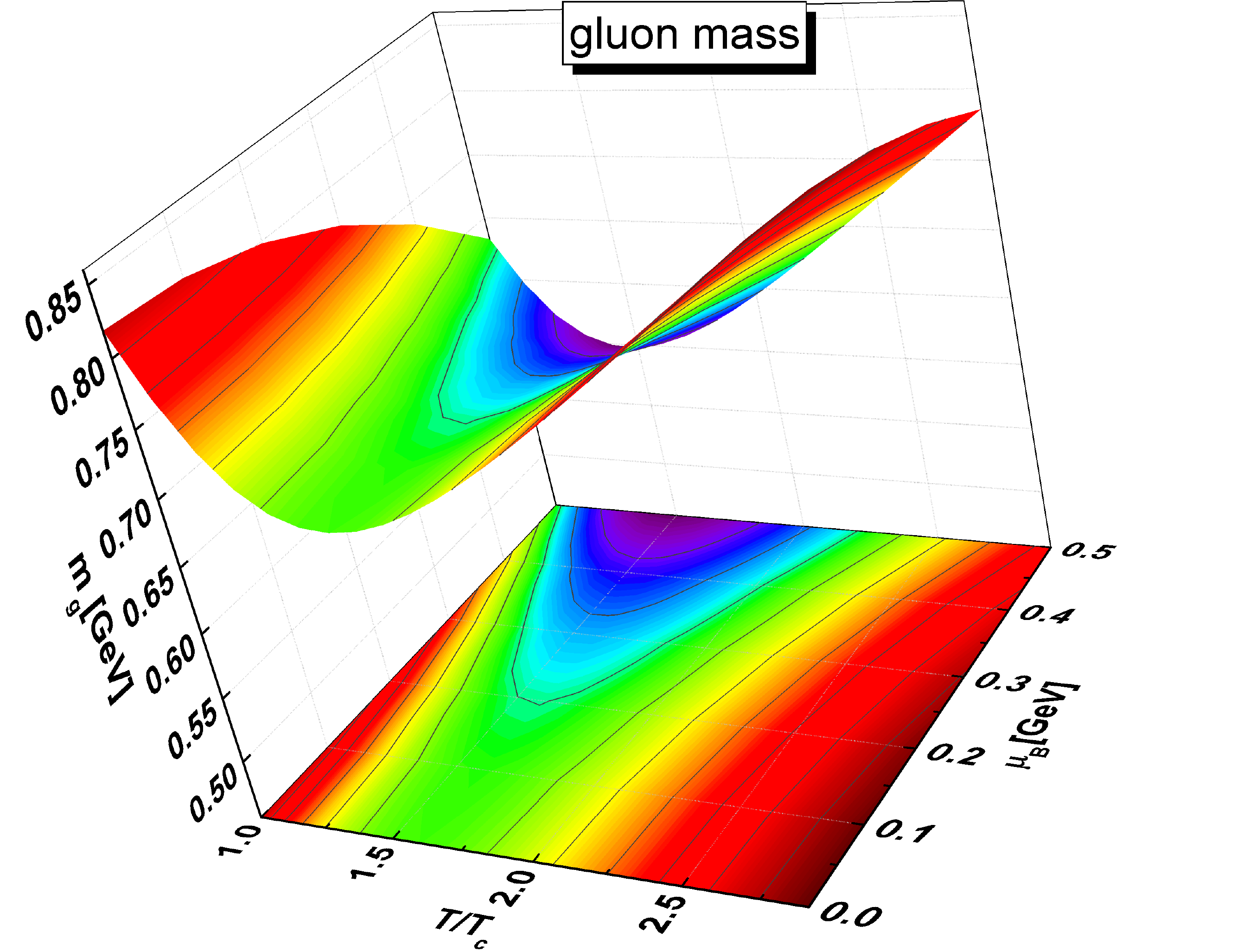}}
\centerline{ \includegraphics[width=75mm]{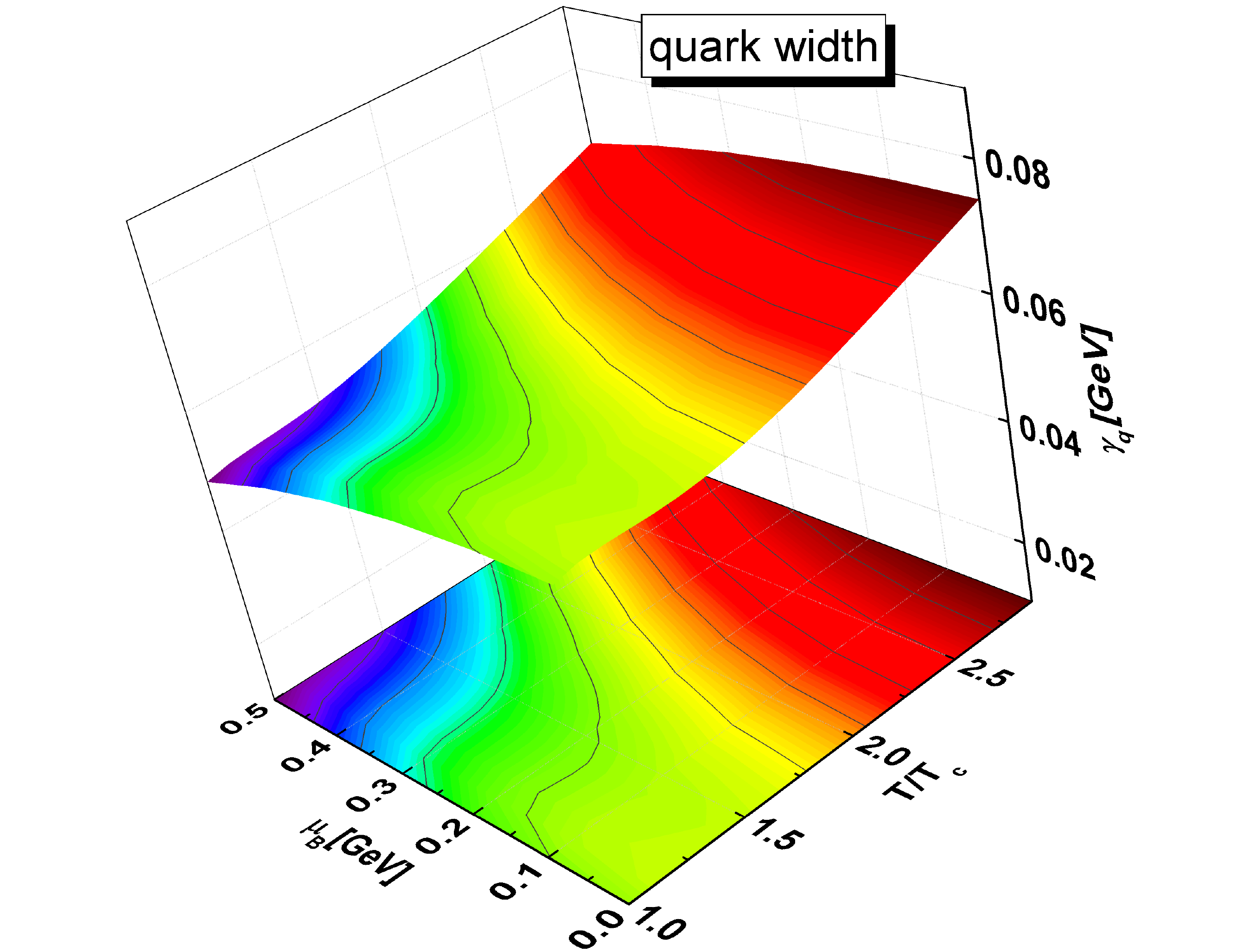}
\includegraphics[width=75mm]{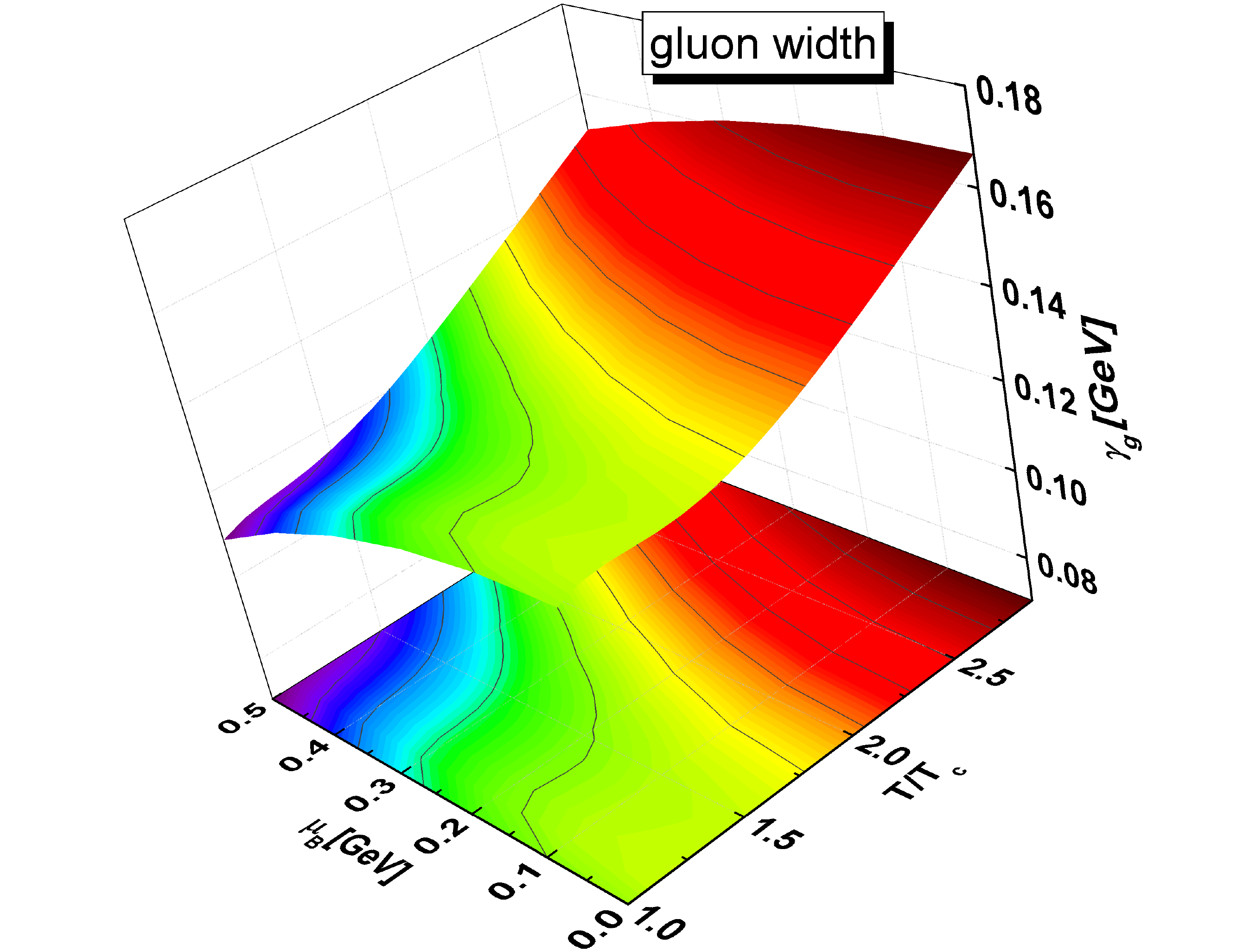}}

\caption{The effective quark (\textbf{left}) and gluon (\textbf{right}) masses $M$
(\textbf{upper row}) and widths $\gamma$ (\textbf{lower row}) from the actual DQPM 
 as a function of the temperature $T$ and baryon chemical potential $\mu_B$ \cite{Olga2020}. }
\label{fig1}
\end{figure}

The actual parameters in Eq. (\ref{eq:rho}),  i.e. the gluon
mass $M_g$ and width $\gamma_g$---employed as input in the present PHSD
calculations---as well as the quark mass $M_q$ and width
$\gamma_q$, are depicted in Fig. \ref{fig1} as a function of the
 temperature $T$ and  baryon chemical potential $\mu_B$ (from Ref. \cite{Olga2020}). These values for the masses and widths
have been fixed by fitting the lattice QCD results from Ref.~\cite{Borsanyi:2012cr,Borsanyi:2013bia} in thermodynamic equilibrium.
One can see that the masses of quarks and gluons decrease with
increasing $\mu_B$, and a similar trend holds for the widths of these partons.

A scalar mean-field $U_s(\rho_s)$ for quarks and antiquarks can be
defined by the derivative of the potential energy density with respect to the
scalar density $\rho_s(T,\mu_B)$,
\begin{equation} \label{uss}
U_s(\rho_s) = \frac{d V_p(\rho_s)}{d \rho_s} ,
\end{equation}
which is evaluated numerically within the DQPM. Here, the potential energy density
is defined by
\begin{equation} \label{Vp}
V_p(T,\mu_B) = T^{00}_{g-}(T,\mu_B) + T^{00}_{q-}(T,\mu_B) + T^{00}_{{\bar q}-}(T,\mu_B),
\end{equation}
where the different contributions $T^{00}_{j-}$ correspond to the
space-like part of the energy-momentum tensor component $T^{00}_{j}$
of parton $j = g, q, \bar{q}$ (cf. Section~3 in Ref.
\cite{Cassing:2007nb}). The scalar mean-field $U_s(\rho_s)$ for quarks
and antiquarks is repulsive as a function of the parton scalar
density $\rho_s$ and shows that the scalar mean-field potential is in the
order of a few GeV for $\rho_s > 10$ fm$^{-3}$. The mean-field potential
(\ref{uss}) is employed in the PHSD transport calculations and
determines the force on a partonic quasiparticle $j$, i.e.,~$ \sim
M_j/E_j \nabla U_s(x) = M_j/E_j \ d U_s/d \rho_ s \ \nabla
\rho_s(x)$, where the scalar density $\rho_s(x)$ is determined
numerically on a space-time~grid in PHSD.

Furthermore, a two-body interaction strength can be extracted from
the DQPM as well from the quasiparticle widths in line with Ref.~\cite{Peshier:2005pp}. On the partonic side, the following elastic and inelastic
interactions are included in the latest version of PHSD (v. 5.0)
$qq \leftrightarrow qq$, $\bar{q} \bar{q} \leftrightarrow \bar{q}\bar{q}$, $gg \leftrightarrow gg$,
$gg \leftrightarrow g$, $q\bar{q} \leftrightarrow g$, $q g \leftrightarrow q g$,
$g \bar{q} \leftrightarrow g \bar{q}$  exploiting
'detailed-balance' with cross sections calculated from the leading order Feynman diagrams employing the effective propagators and couplings $g^2(T/T_c)$ from the DQPM~\cite{Moreau:2019vhw}. In~Ref.~\cite{Moreau:2019vhw}, the differential and total off-shell
cross sections have been evaluated as a function of the invariant energy of
colliding off-shell partons $\sqrt{s}$ for each $T$, $\mu_B$.
We recall that in the previous PHSD studies (using version 4.0 and below) the cross sections depend only on $T$ (cf. the detailed evaluation in Ref.~\cite{Ozvenchuk13}).
When implementing the differential cross sections and parton masses into the PHSD5.0
approach, one has to specify the 'Lagrange parameters' $T$ and $\mu_B$ in each
computational cell in space-time. This has been done by employing the lattice equation
of state and a diagonalization of the energy-momentum tensor from PHSD as described in Ref. \cite{Moreau:2019vhw}.

The transition from partonic to hadronic degrees-of-freedom (and vice~versa)  is described by
covariant transition rates for the fusion of quark--antiquark pairs
or three quarks (antiquarks), respectively, obeying flavor
current--conservation, color neutrality as well as energy--momentum
conservation~\cite{Cassing:2009vt}. Since the dynamical quarks and
antiquarks become very massive close to the phase transition, the formed resonant 'prehadronic' color-dipole states ($q\bar{q}$ or
$qqq$) are of high invariant mass, too, and sequentially decay to
the groundstate meson and baryon octets, thus increasing the total
entropy during hadronization.

On the hadronic side, PHSD includes explicitly the baryon octet and
decouplet, the~$0^-$- and $1^-$-meson nonets as well as selected
higher resonances as in the Hadron--String--Dynamics (HSD)
approach~\cite{Cassing:1999es}.  Note that PHSD and HSD
(without explicit partonic degrees-of-freedom) merge at low energy
density, in particular below the local critical energy density
$\varepsilon_c\approx$ 0.5~GeV/fm$^{3}$ as extracted from the lQCD results in Ref.~\cite{Borsanyi:2012cr,Borsanyi:2013bia}.

\section{Transport~Coefficients}

The transport properties of the QGP close to equilibrium can be characterized by
various transport coefficients.
The shear viscosity $\eta$ and bulk viscosity $\zeta$ describe the fluid's
dissipative corrections at leading order. Both coefficients are generally expected
to depend on the temperature $T$ and baryon chemical potential $\mu_B$.
In the hydrodynamic equations, the viscosities appear as dimensionless ratios,
$\eta/s$ and $\zeta/s$, where $s$ is the fluid entropy density.
Such specific viscosities are more meaningful than the unscaled $\eta$ and $\zeta$
values because they describe the magnitude of stresses inside
the medium relative to its natural scale.

In our recent studies \cite{Moreau:2019vhw,Soloveva:2019xph,Soloveva:2019doq}, we have
investigated the transport properties of the QGP in the $(T,\mu_B)$ plane based on the DQPM.
One way to evaluate the viscosity coefficients of partonic matter is the Kubo formalism~\cite{Kubo,Aarts:2002,Fraile:2006,Lang12}, which was used to calculate the viscosities for a previous version of the DQPM within the PHSD transport approach in a box with periodic boundary conditions in Ref.~\cite{Ozvenchuk13:kubo} as well as in the more recent study with
the DQPM model in Refs.~\cite{Moreau:2019vhw,Soloveva:2019xph}.
Another way to calculate transport coefficients (explored also in~\cite{Moreau:2019vhw,Soloveva:2019xph}) is to use the relaxation--time approximation (RTA)
as incorporated in Refs. \cite{Hosoya:RTA,Chakraborty11,Kapusta,SGavin}.

The shear viscosity based on the RTA (cf.~\cite{Sasaki:2008fg}) reads as:
\begin{equation}
\eta^{{RTA}}(T,\mu_q)  = \frac{1}{15T} \sum_{i=q,\bar{q},g} \int \frac{d^3p}{(2\pi)^3} \frac{\mathbf{p}^4}{E_i^2}     \tau_i(\mathbf{p},T,\mu)
\  d_i  (1 \pm f_i) f_i , \label{eta_on}
\end{equation}
where  $d_q = 2N_c = 6$ and $d_g = 2(N_c^2-1) = 16$ are degeneracy factors for spin and color in case of quarks and  gluons , whereas $\tau_i$ are relaxation times for particles $i$.
Equation~(\ref{eta_on}) includes the Bose enhancement and Pauli-blocking factors, respectively, which are taken into account in the actual calculations. The pole energy is $E_i^2 = p^2 + M_i^2$, where $M_i$ is the pole mass from the DQPM. The notation $\sum_{j=q,\bar{q},g}$ includes the contribution from all possible partons which in our case are the gluons and the (anti-)quarks of three different flavors ($u,d,s$).

We consider two cases for the relaxation time for quarks and gluons:
(1) $\tau_i(\mathbf{p},T,\mu) =1/ \Gamma_i(\mathbf{p},T,\mu)$
and (2) $\tau_i(T,\mu) =1/ 2\gamma_i(T,\mu)$,
where $\Gamma_i(\mathbf{p},T,\mu)$ is the parton interaction rate, calculated microscopically from the collision integral using the differential cross sections
for parton scattering from Ref. \cite{Moreau:2019vhw}, while $\gamma_i(T,\mu)$ are the width parameters in the parton propagators (\ref{eq:rho}).

\begin{figure}[h]
{\includegraphics[width=7.8cm]{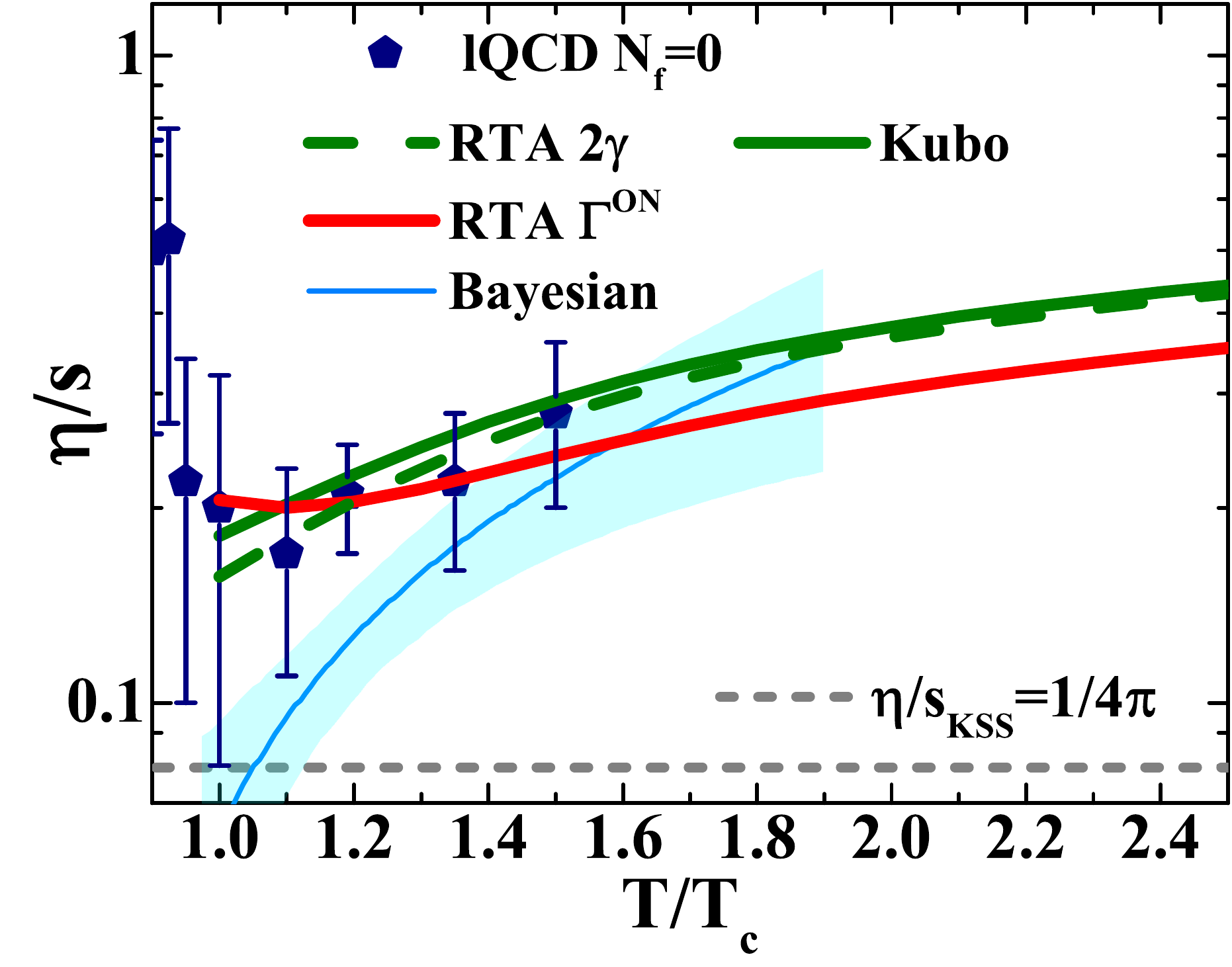}}
{\includegraphics[width=8.4cm]{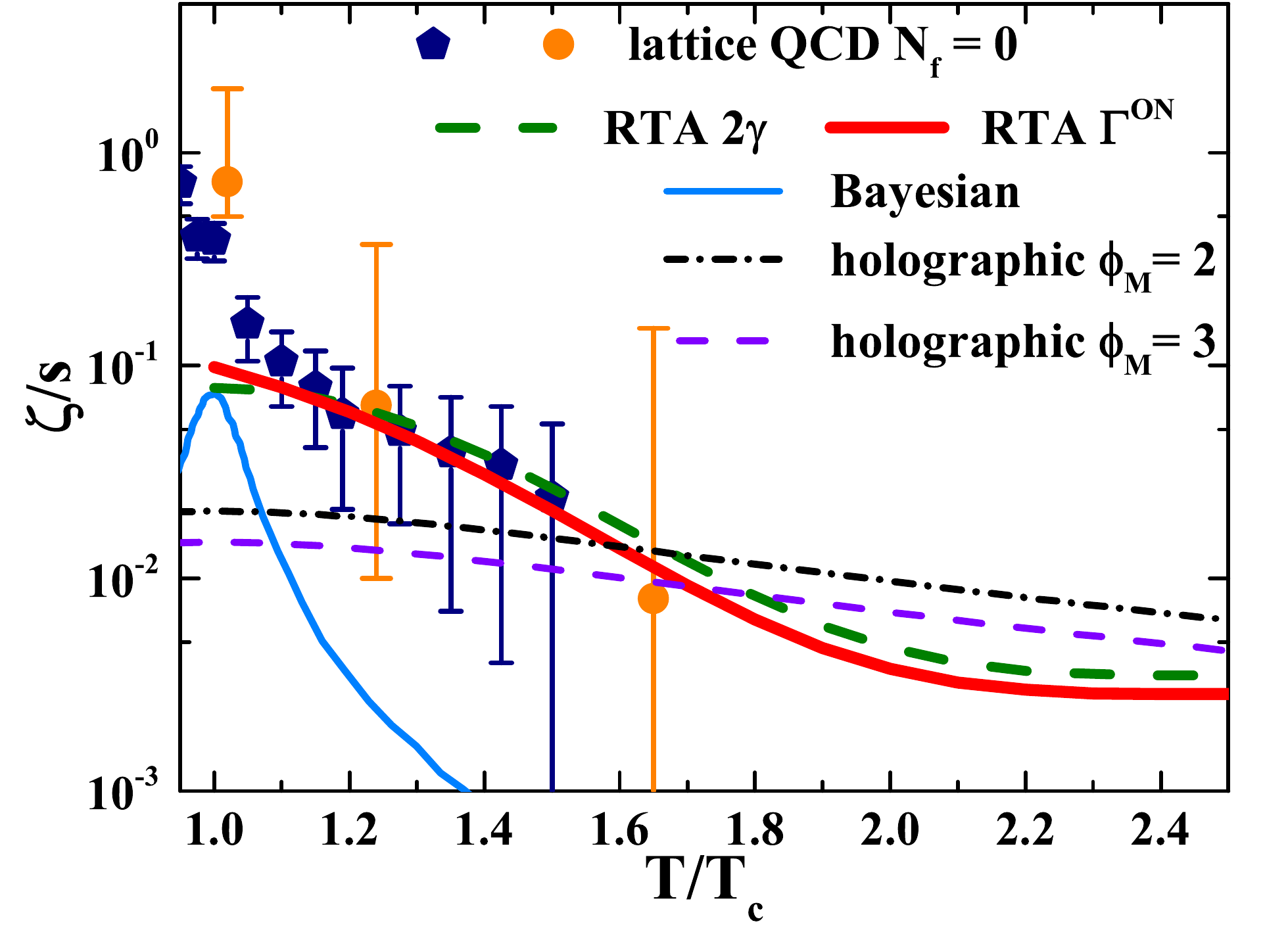}}
\caption{ (\textbf{lhs}): the ratio of shear viscosity to entropy density $\eta/s$ as a function of
the scaled temperature $T/T_c$ for $\mu_B = 0$ calculated within the Kubo formalism
(green solid line) and the RTA approach with the interaction rate $\Gamma^{\rm{on}}$
(red solid line) and the DQPM width ${2\gamma}$ (dashed green line).
 The dashed gray line demonstrates the Kovtun--Son--Starinets bound~\cite{Kovtun:2004de} $(\eta/s)_{\rm{KSS}} = 1/(4\pi)$, and the symbols show lQCD data for pure SU(3) gauge theory taken from Ref.~\cite{Astrakhantsev:2017} (pentagons). The solid blue line shows the results from a Bayesian analysis of experimental heavy-ion data from Ref.~\cite{Bernhard:2016tnd}.
 (\textbf{rhs}): the ratio of the bulk viscosity to entropy density $\zeta/s$ as a function of  the scaled temperature $T/T_c$ for $\mu_B=0$ calculated using the RTA approach with
the on-shell interaction rate $\Gamma^{\rm{on}}$ (red solid line)
and the DQPM width ${2\gamma}$ (dashed green line). The symbols correspond to the lQCD data for pure SU(3) gauge theory taken from Refs.~\cite{Astrakhantsev:2018oue} (pentagons) and~\cite{Meyer} (circles). The solid blue line shows the results from a Bayesian analysis of experimental heavy-ion data from Ref.~\cite{Bernhard:2016tnd}.
The dot-dashed and dashed lines correspond to the results from the non-conformal holographic model for $\phi_M=2$ and 3, correspondingly, from~Ref.~\cite{Maximilian}.}
\label{fig_eta_zeta}
\end{figure}

In the left part  of Fig. \ref{fig_eta_zeta} we show
the ratio of the shear viscosity to entropy density as a function of the scaled temperature
$T/T_c$ for $\mu_B = 0$ calculated within the Kubo formalism (green solid line) and RTA approach
with the interaction rate $\Gamma^{\rm{on}}$ (red solid line) and the DQPM width ${2\gamma}$ (dashed green line).
The RTA approximation (\ref{eta_on}) of the shear viscosity with the DQPM width
${2\gamma}$ and with the interaction rate $\Gamma^{\rm{on}}$ are quite close to each other
at $\mu_B=0$ and also very close to the result from the Kubo formalism~\cite{Moreau:2019vhw}  indicating that the quasiparticle limit ($\gamma \ll M$) holds in the DQPM.

The ratio $\eta/s$ increases with the scaled temperature.
The actual values for the ratio $\eta/s$  are in a good agreement with the gluodynamic
lattice QCD calculations at $\mu_B=0$ from Ref.~\cite{Astrakhantsev:2017}.
Moreover, our DQPM results are in qualitative agreement with the results
from a Bayesian analysis of experimental heavy-ion data from Ref.~\cite{Bernhard:2016tnd}.
We mention that the DQPM result differs from the recent calculations for the
shear viscosity at $\mu_B=0$ in the quasiparticle model in Ref.~\cite{Mykhaylova:2019}
where the width of quasiparticles is not considered, which leads to a higher
value for the $\eta/s$ ratio. This shows the sensitivity of this ratio to the modelling of
partonic interactions and the properties of partons in the hot QGP medium.
We recall also that in Refs.~\cite{Moreau:2019vhw,Soloveva:2019xph,Soloveva:2019doq} we have found that the ratio $\eta/s$ shows a very weak dependence on $\mu_B$ and has a similar behavior as a function of temperature for all $\mu_B \le 400$ MeV.

The expression for the bulk viscosity in the partonic phase - derived within the RTA -  reads (following  Ref.~\cite{Kapusta})
\begin{equation}
\zeta^{{RTA}}(T,\mu)= \frac{1}{9T} \sum_{i=q,\bar{q},g}\int \frac{d^3p}{(2\pi)^3}
 \tau_i(\mathbf{p},T,\mu)
\ \frac{ d_i  (1 \pm f_i) f_i  }{E_i^2}\left(\mathbf{p}^2-3c_s^2\left(E_i^2-T^2\frac{dm_i^2}{dT^2}\right)\right)^2 ,
  \label{zeta_on} \end{equation}
where $c_s^2$ is the speed of sound squared, and $\frac{dm_i^2}{dT^2}$ is the DQPM parton mass derivative which becomes large close to the critical temperature $T_c$.

On the right side of Fig. \ref{fig_eta_zeta} we show the ratio of the bulk viscosity to
entropy density $\zeta/s$ as a function of  the scaled temperature $T/T_c$ for $\mu_B=0$
calculated by the RTA approach with the interaction rate $\Gamma^{\rm{on}}$
(red solid line) and the DQPM width ${2\gamma}$ (dashed green line).
The symbols correspond to the lQCD data for pure SU(3) gauge theory taken from
Refs.~\cite{Astrakhantsev:2018oue} (pentagons) and~\cite{Meyer} (circles).
The solid blue line shows the results from a Bayesian analysis of experimental
heavy-ion data from Ref.~\cite{Bernhard:2016tnd}.
The dot-dashed and dashed lines correspond to the results from the non-conformal holographic
model~\cite{Maximilian} for $\phi_M=2$ and 3, correspondingly, where $\phi_M$
is the model parameter which characterizes the non-conformal features of the model.
We find that the DQPM result for $\zeta/s$ is in good agreement with the lattice QCD results and shows a rise closer to
$T_c$ contrary to the holographic results, which show practically a constant behavior independent of  model parameters. This rise is attributed to the increase of the partonic
mass closer to  $T_C$ as shown in Figure~\ref{fig1}, thus the mass derivative term
in Eq. (\ref{zeta_on}) also grows. The~Bayesian result also shows a peak near $T_C$;
however, the~ratio drops to zero while lQCD data indicate a positive $\zeta/s$
as found also in the DQPM.
The explicit $\mu_B$ dependence of $\zeta/s$ has been investigated within the DQPM in Refs. \cite{Moreau:2019vhw,Soloveva:2019xph,Soloveva:2019doq}, where it has been shown that
it is rather weak for $\mu_B \le 400$ MeV. 

As follows from hydrodynamical calculations, the results for the flow harmonics $v_n$
are sensitive to the transport coefficients~\cite{Bernhard:2016tnd,Marty:NJL13,Romatschke,Song:Heinz}.
Thus, there are hopes to observe a $\mu_B$ sensitivity of $v_1, v_2$. We will study this sensitivity in the following Section.

\section{Heavy-Ion~Collisions}

In our recent study~\cite{Moreau:2019vhw} we have investigated the sensitivity
of 'bulk' observables such as rapidity and transverse momentum distributions
of different hadrons produced in heavy-ion collisions from AGS to top RHIC energies on the details of the QGP interactions and the properties of the partonic degrees-of-freedom.
For that, we have considered the following three cases:

(1) {\bf `PHSD4.0'}: the masses and widths of quarks and gluons depend only on $T$.
The cross sections for partonic interactions depend only on $T$ as evaluated by
the `box' calculations in Ref.~\cite{Ozvenchuk13} in order to merge the QGP interaction
rates from all possible partonic channels to the total temperature dependent widths $2 \gamma_i$ 
of the DQPM propagator. This has been used in the PHSD code (version 4.0 or below) for extended studies of many hadronic observables in p+A and A+A collisions
at different energies~\cite{Linnyk:2015rco,Cassing:2008sv,Cassing:2008nn,Cassing:2009vt,Bratkovskaya:2011wp,Konchakovski:2011qa} with good success.

(2) {\bf `PHSD5.0 - $\mu_B=0$'}: the masses and widths of quarks and gluons depend only
on $T$; however, the~differential and total partonic cross sections are obtained by calculations of the leading order Feynman 
diagrams employing the effective propagators and couplings $g^2(T/T_c)$ from the DQPM
at $\mu_B=0$ \cite{Moreau:2019vhw}. Thus, the cross sections depend explicitly on the
invariant energy of the colliding partons $\sqrt{s}$ and on $T$. This is realized in the PHSD5.0 by setting $\mu_B=0$ (cf.~\cite{Moreau:2019vhw}).

(3) {\bf `PHSD5.0 - $\mu_B$'}: the masses and widths of quarks and gluons depend on $T$
and $\mu_B$ explicitly; the differential and total partonic cross sections are obtained by calculations of the leading order Feynman diagrams from the DQPM and explicitly depend on invariant energy $\sqrt{s}$, temperature $T$ and baryon chemical potential $\mu_B$.
This is realized in the full version of PHSD5.0 (cf.~\cite{Moreau:2019vhw}).

The comparison of the 'bulk' observables for A+A collisions within the three cases of PHSD in Ref.~\cite{Moreau:2019vhw}
has illuminated that they show a very low sensitivity to the $\mu_B$ dependences of parton
properties (masses and widths) and their interaction cross sections such that the results
from PHSD5.0 with and without $\mu_B$ were very close to each other.
Only in the case of kaons,  antiprotons $\bar{p}$ and antihyperons $\bar{\Lambda} + \bar{\Sigma}^0$, a small difference between PHSD4.0 and PHSD5.0 could be seen at
top SPS and top RHIC energies. A similar trend has been found for very asymmetric collisions of C+Au: a small sensitivity to the partonic scatterings was found in the kaon and antibaryon rapidity distributions, too.
This can be understood as following: at high energies such as top RHIC where the QGP volume
is very large in central collisions, the~$\mu_B$ is very low, while, when
decreasing the bombarding energy (or increasing $\mu_B$) 
the fraction of the QGP is decreasing such that the final observables
are dominated by the hadronic phase, i.e.,~the probability for
the hadrons created at the QGP hadronization to rescatter, decay, or be absorbed
in hadronic matter increases strongly; as a result the sensitivity to the properties of the QGP is washed out.

\subsection{Directed~Flow}

\begin{figure}[t]
\begin{center}
\includegraphics[width=12cm]{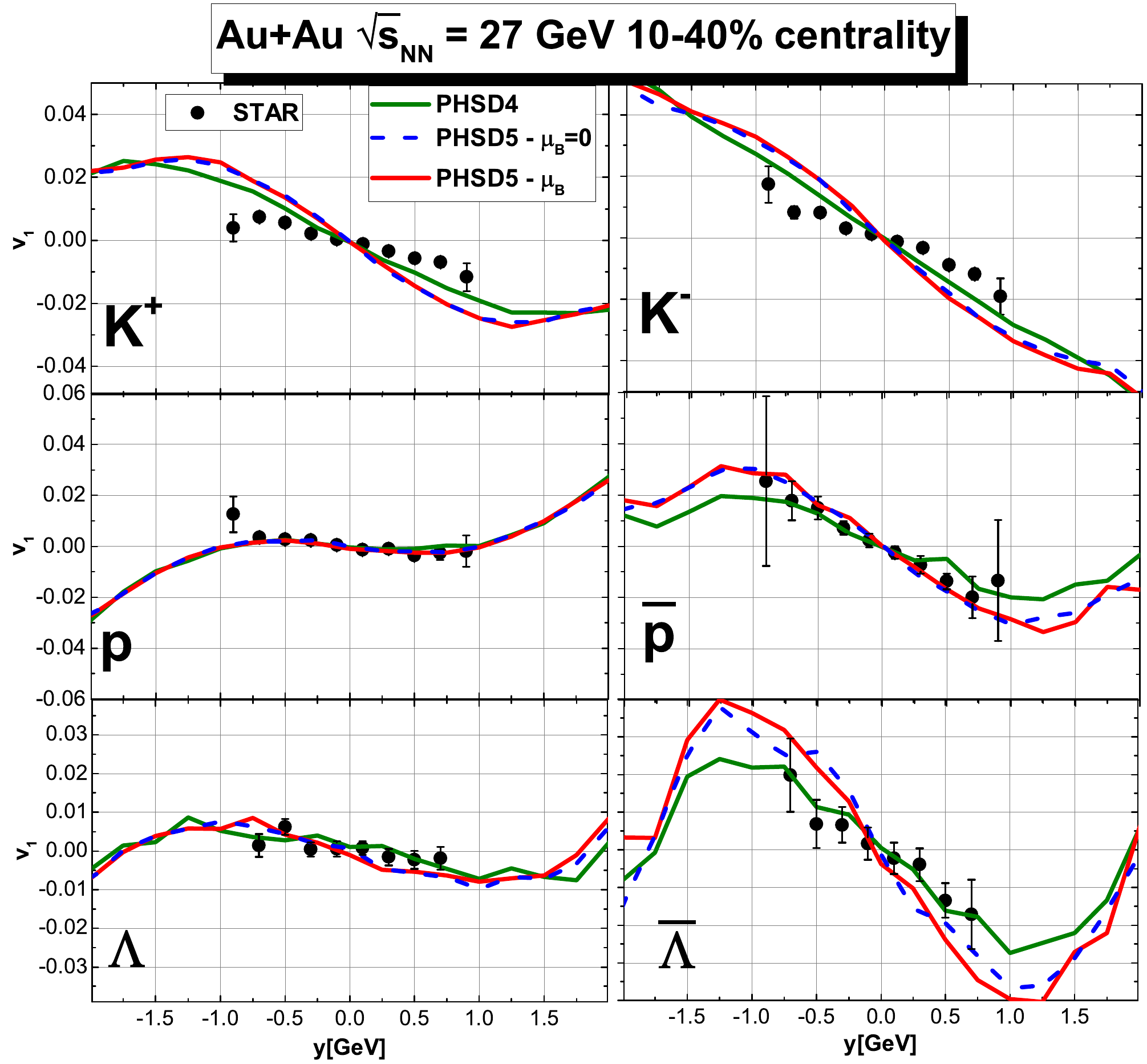}
\end{center}
\caption{Directed flow of identified hadrons as a function of rapidity for Au+Au collisions at $\sqrt{s_{NN}}$ = 27 GeV for PHSD4.0 (green lines), PHSD5.0 with partonic cross sections and parton masses calculated for $\mu_B$ = 0 (blue dashed lines) and with cross sections and parton masses evaluated at the actual chemical potential $\mu_B$ in each individual space-time cell (red lines) in comparison to the experimental data of the STAR Collaboration~\cite{STARBESv1}.}
\label{v1y27GeV}
\end{figure}

We here test the traces of $\mu_B$ dependences of the QGP interaction cross sections in collective observables such as the directed flow $v_1$ considering again the three cases for the PHSD as discussed above. We directly report about the actual PHSD results. 
Fig. \ref{v1y27GeV} depicts the directed flow $v_1$ of identified hadrons
($K^\pm, p, \bar p, \Lambda+\Sigma^0, \bar\Lambda + \bar\Sigma^0$) versus
rapidity for Au+Au collisions at $\sqrt{s_{NN}}$ = 27 GeV.
One can see a good agreement between PHSD results and experimental data from the STAR collaboration~\cite{STARBESv1}.
However, the different versions of PHSD for the $v_1$ coefficients show a quite
similar behavior; only antihyperons indicate a slightly different flow.
This supports again the finding that strangeness, and in particular anti-strange
hyperons, are the most sensitive probes for the QGP~properties. Surprizingly, PHSD4.0 performs better for the directed flows of hadrons than PHSD5.0 with the microscopic differential partonic cross sections.

\subsection{Elliptic~Flow}
Hydrodynamic simulations \cite{Romatschke,Song:Heinz} and 
the Bayesian analysis~\cite{Bernhard:2016tnd} indicate that the elliptic flow $v_2$ is sensitive
to the transport properties of the QGP as characterized by transport coefficients such as shear $\eta$ and bulk $\zeta$ viscosities.
In this subsection we present the results for the elliptic flow of charged hadrons
from HICs within the PHSD5.0 with and without $\mu_B$ dependence and compare the results with those from PHSD4.0 as before.

\begin{figure}[h]
\centering
\includegraphics[width=12cm]{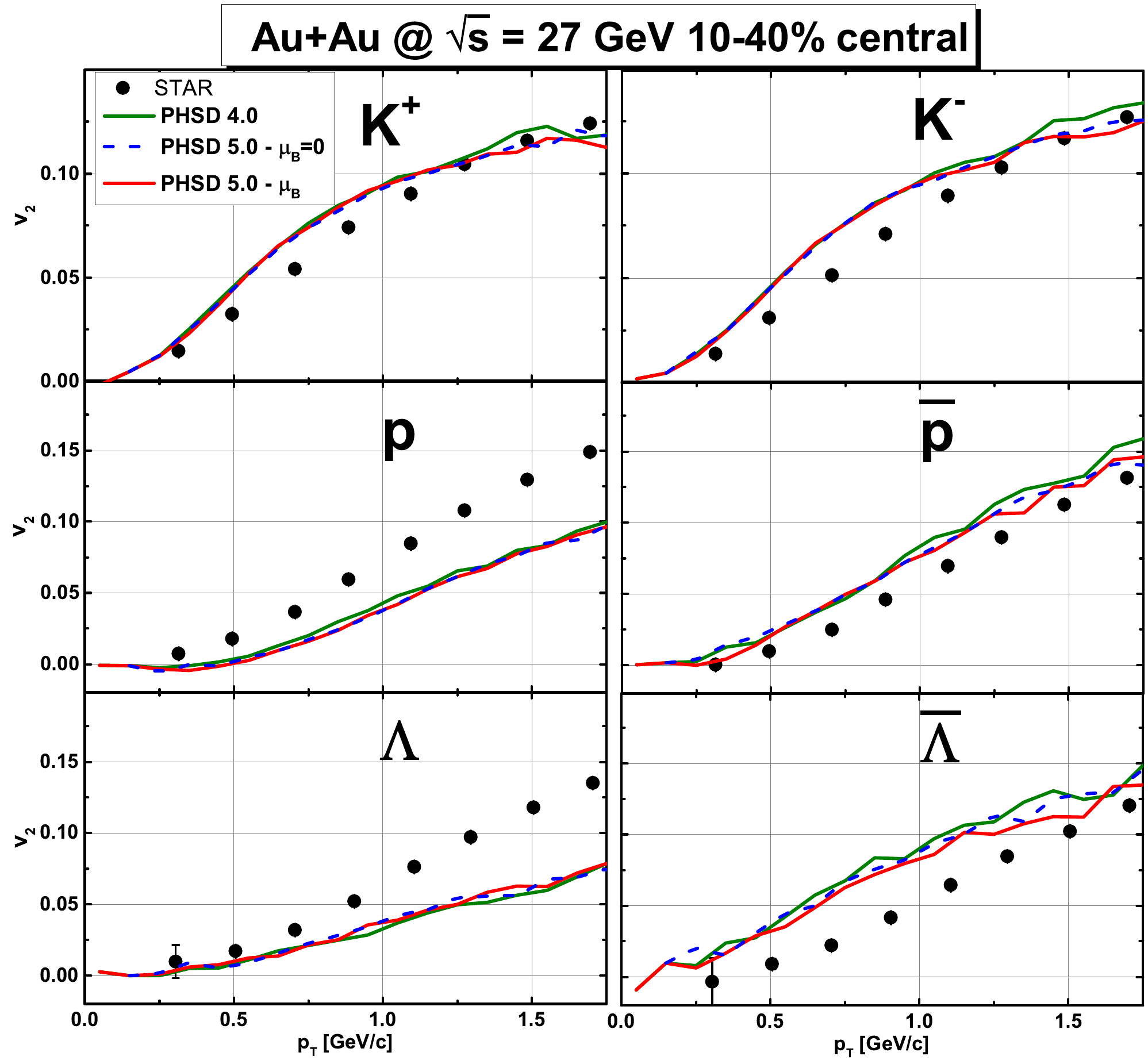}
\caption{Elliptic flow of identified hadrons ($K^\pm, p, \bar p, \Lambda+\Sigma^0, \bar\Lambda + \bar\Sigma^0$) as a function of $p_T$ for Au+Au collisions at $\sqrt{s_{NN}}$ = 27~GeV  for PHSD4.0 (green lines), PHSD5.0 with partonic cross sections and parton masses calculated for $\mu_B$ = 0 (blue dashed lines) and with cross sections and parton masses evaluated at the actual chemical potential $\mu_B$ in each individual space-time cell (red lines)
in comparison to the experimental data of the STAR Collaboration~\cite{STARBESv2}.}
\label{v2y27GeV}
\end{figure}

In Fig. \ref{v2y27GeV} we display the elliptic flow $v_2$ of identified hadrons ($K^\pm, p, \bar p, \Lambda+\Sigma^0, \bar\Lambda + \bar\Sigma^0$) as a function of  $p_T$ at
$\sqrt{s_{NN}}$ = 27 GeV  for PHSD4.0 (green lines), PHSD5.0 with partonic cross sections and parton masses calculated for $\mu_B$ = 0 (blue dashed lines) and with cross sections and parton masses evaluated at the actual chemical potential $\mu_B$ in each individual space-time cell (red lines) in comparison to the experimental data of the STAR Collaboration~\cite{STARBESv2}.
Similar to the directed flow shown in Figure~\ref{v1y27GeV}, the~elliptic flow from all
three cases for PHSD shows a rather similar behavior;  the differences are very small (within the statistics achieved here). Only antiprotons and antihyperons show a small decrease of
$v_2$ at larger $p_T$ for PHSD5.0 compared to PHSD4.0, which can be attributed
to the explicit $\sqrt{s}$-dependence and different angular distribution of partonic cross sections in the PHSD5.0. We note that the underestimation of $v_2$ for protons
and $\Lambda$'s might be attributed to the details of the hadronic vector potentials involved
in this calculations which seem to underestimate the baryon repulsion.

\section{Conclusions}

In this contribution we have reported on the influence of the baryon chemical potential $\mu_B$
on the properties of the QGP in equilibrium as well as the QGP created in heavy-ion collisions initially also far from equilibrium.
For the description of the QGP we have employed the extended 
Dynamical QuasiParticle Model (DQPM) that is matched to reproduce the lQCD equation-of-state (versus temperature $T$) at zero and finite baryon chemical potential $\mu_B$.
The DQPM results for transport coefficients, such as shear viscosity $\eta$
and bulk viscosity $\zeta$, have been compared to available lQCD data, the non-conformal holographic model at $\mu_B=0$
and with results from a Bayesian analysis of experimental heavy-ion data.
We find that the ratios $\eta/s$ and $\zeta/s$ from the DQPM agree very well with
the lQCD results from Ref.~\cite{Astrakhantsev:2018oue} and show a similar behavior
as the ratio obtained from a Bayesian fit~\cite{Bernhard:2016tnd}. As found previously in Refs. \cite{Moreau:2019vhw,Soloveva:2019xph} the transport coefficients show only a mild dependence on $\mu_B$.

Our study of the non-equilibrium QGP---as created
in relativistic heavy-ion collisions---has been performed within the extended Parton--Hadron--String Dynamics (PHSD)
transport approach ~\cite{Moreau:2019vhw} in which i) the masses and widths of quarks and gluons depend on $T$
and $\mu_B$ explicitly; ii) the partonic interaction cross sections are obtained
by the leading order Feynman diagrams from the DQPM effective propagators and couplings  and explicitly depend on
the invariant energy $\sqrt{s}$, temperature $T$ and baryon chemical potential $\mu_B$.
This extension is realized in the  version PHSD5.0 \cite{Moreau:2019vhw}.
In order to investigate the traces of the $\mu_B$ dependence of the QGP in observables,
the results of PHSD5.0 (with $\mu_B$ dependences) have been compared to the
results of PHSD5.0 for $\mu_B=0$ as well as with PHSD4.0, where
the masses/widths of quarks and gluons as well as their interaction cross sections
depend only on $T$ (as specified in Ref. \cite{Ozvenchuk13}). 

Since the sensitivity (w.r.t. the $\mu_B$-dependence) of hadronic rapidity and $p_T$ distributions of identified hadrons for symmetric and asymmetric heavy-ion collisions from AGS to RHIC energies was found to be very low \cite{Moreau:2019vhw,Soloveva:2019xph} we have focused on the related sensitivity of the collective  flow of hadrons. As a characteristic example we have shown  
(i) the directed flow $v_1$ of identified hadrons  and
(ii) the elliptic flow  $v_2$ of identified hadrons for $Au+Au$ collisions  at the invariant
energy $\sqrt{s_{NN}}=27$  GeV.
We find only very small differences between the results from PHSD4.0 and from PHSD5.0 on the hadronic flow observables at high as well as at intermediate energies.  This is related to the fact that at high energies, where the matter is dominated by the QGP,
one probes only a small baryon chemical potential in central collisions at midrapidity,
while with decreasing energy (and larger $\mu_B$) the fraction of the QGP drops rapidly, such that in total the final observables are
dominated by the hadronic interactions and thus the information about the partonic properties and scatterings is washed out.
We have shown that the mild $\mu_B$ dependence of QGP interactions is more
pronounced in observables for strange hadrons (kaons and especially
anti-strange hyperons) which provides an experimental hint for the search of  $\mu_B$ traces of the QGP for experiments at the future FAIR/NICA facilities and the BESII program at RHIC.

\section*{Acknowledgements}
The authors acknowledge inspiring discussions with J\"org Aichelin,
 Vadim Kolesnikov,  Ilya  Selyuzhenkov and Arkadiy Taranenko. Furthermore, they thank Maximillian Attems
for providing the results from Ref. \cite{Mykhaylova:2019} in data form.
This work was supported by the LOEWE center "HIC for FAIR", the HGS-HIRe for FAIR and the COST Action THOR, CA15213.
LO is supported by the Alexander von Humboldt-Stiftung.
Furthermore, PM and EB acknowledge support by DFG through the grant CRC-TR 211 'Strong-interaction matter under extreme conditions'. The computational resources have been provided by the LOEWE-Center for Scientific Computing.\\

\end{document}